\documentclass{llncs}
\usepackage{amsmath}
\usepackage[english]{babel}
\usepackage{graphicx}
\usepackage[bookmarks=false,colorlinks=true,linkcolor={blue},pdfstartview={XYZ null null 1.22}]{hyperref} 

\title{Integrating the Probabilistic Model BM25/BM25F into Lucene.}
\author{Joaqu\'in P\'erez-Iglesias\inst{1}, Jos\'e R. P\'erez-Ag\"uera\inst{2}, V\'ictor Fresno\inst{1} and\\ Yuval Z. Feinstein\inst{3}}
\institute{NLP\&IR Group, Universidad Nacional de Educaci\'on a Distancia, Spain
\and
 University of North Carolina at Chapel Hill,
 USA\\
\and Answers Corporation, Jerusalem 91481, Israel\\
\email{joaquin.perez@lsi.uned.es, jaguera@email.unc.edu, vfresno@lsi.uned.es, yuvalf@answers.com}
}

\begin{document}
\maketitle
\begin{abstract}
This document describes the BM25 and BM25F implementation using the Lucene Java
Framework. The implementation described here can be downloaded from
\cite{jperezi08}. Both models have stood out at TREC by their performance and are considered as state-of-the-art in the IR community. BM25 is applied to retrieval on plain text documents, that is for documents that do not contain fields, while BM25F is applied to documents with structure. 
\end{abstract}

\section*{Introduction}
Apache Lucene is a high-performance and full-featured text search engine library written entirely in Java. It is a technology suitable for nearly any application that requires full-text search. Lucene is scalable and offers high-performance indexing, and has become one of the most used search engine libraries in both academia and industry \cite{Lucene09}. 

Lucene ranking function, the core of any search engine applied to determine how relevant a document is to a given query, is built on a combination of the Vector Space Model (VSM) and the Boolean model of Information Retrieval.  The main idea behind Lucene approach is the more times a query term appears in a document relative to the number of times the term appears in the whole collection, the more relevant that document will be to the query \cite{Lucene09}. Lucene uses also the Boolean model to first narrow down the documents that need to be scored based on the use of boolean logic in the query specification. 

In this paper, the implementation of BM25 probabilistic model and its extension for semi-structured IR, BM25F, is described in detail. 

One of the main Lucene's constraints to be widely used by IR community is the lack of different retrieval models implementations. Our goal with this work is to offer to IR community a more advanced ranking model which can be compared with other IR software, like Terrier, Lemur, CLAIRlib or Xapian.

\section{Motivation}

There exists previous implementations of alternative Information Retrieval Models for Lucene. The most representative case of that is the Language Model implementation\footnote{http://ilps.science.uva.nl/resources/lm-lucene} from Intelligent Systems Lab Amsterdam. Another example is described at \cite{Doron07} where Lucene is compared with Juru system. In this case Lucene document length normalization is changed in order to improve the Lucene ranking function performance.

BM25 has been widely use by IR researchers and engineers to improve search engine relevance, so from our point of view, a BM25/BM25F implementation for Lucene becomes necessary to make Lucene more popular for IR community.


\section*{Included Models}
The developed models are based in the information that can be found at \cite{Robertson07}. More specifically the
implemented ranking functions are as next: 

\subsection*{BM25}

\begin{equation*}
 R(q,d) = \sum_{t \in q} \frac{occurs_{t}^{d}}{k_1 ((1-b)+b  \frac{l_d}{avl_d}) + occurs_{t}^{d}}
\end{equation*}
where $occurs_t^d$ is the term frequency of $t$ in $d$; $l_d$ is the document $d$ length; $avl_{d}$ is 
the document average length along the collection; $k_1$ is a free parameter usually chosen as 2 
and $ b \in[0,1]$ (usually 0.75). Assigning 0 to $b$ is equivalent to avoid the process of normalisation and therefore the document length will not affect the final score. If $b$ takes 1, we will be carrying out a full length normalisation. 

The classical inverse document frequency is computed as next:
\begin{equation*}
 idf(t) = \log \frac{N - df(t) + 0.5}{df(t) + 0.5}
\end{equation*}
where $N$ is the number of documents in the collection and $df$ is the number of documents where appears the term $t$.

A different version of this formula, as can be found at Wikipedia\footnote{\href{http://en.wikipedia.org/wiki/Probabilistic\_relevance\_model\_(BM25)}{http://en.wikipedia.org/wiki/Probabilistic\_relevance\_model\_(BM25)}}, 
multiplies the obtained bm25 weight by the constant $ (k_1 + 1)$ 
in order to normalize the weight of terms with a frequency equals to 1 that occurs in documents with an average length. 

\subsection*{BM25F}

First we obtain the accumulated weight of a term over all fields as next: 

\begin{equation*}
\displaystyle \textit{weight(t,d)} = \sum_{\textit{c in d}} \frac{occurs_{t,c}^d \cdot boost_c}{((1-b_c)+b_c \cdot \frac{l_c}{avl_c})}
\end{equation*}
where $l_c$ is the field length; $avl_{c}$ is the average length for the field $c$; $b_{c}$ is a constant related 
to the field length, similar to $b$ in BM25 and $boost_c$ is the boost factor applied to field $c$. 

Next, a non-linear saturation $\frac{weight}{k_1 + weight}$, in order to reduce the effect of term frequency to the final score is applied.

\begin{equation}
 \displaystyle \textit{R(q,d)} = \sum_{\textit{t in q}} idf(t) \cdot \frac{weight(t,d)}{k_1 + weight(t,d)}
\end{equation}

$idf(t)$ is computed as in the BM25 case

\begin{equation}
 \displaystyle idf(t)= \log{\frac{N-df(t)+0.5}{df(t)+0.5}}
\end{equation}
where $N$ is the number of documents in the collection and $df$ is the number of documents where appears the term $t$. 

\subsection*{Implementation}
The main goal of this implementation was to integrate the new ranking model into the search Lucene functionalities. 
In order to accomplish this objective a new Query, Weight, and several Scorers were developed. 
The main functionalities are implemented at Scorer level, since the main responsibilities of Query and Weight are to 
prepare the necessary parameters for the Scorers, and create Scorers instances when the search method is invoked. 
More information in the Query-Weight-Scorer model can be found at \href{http://lucene.apache.org/java/2_4_0/scoring.html}.

\subsubsection*{Query}
The execution of a query can be divided in two parts, a boolean filtering and the documents ranking. 
The boolean filtering is carried out by the Scorers ShouldBooleanScorer, MustBooleanScorer and NotBooleanScorer 
depending on the logic operators applied, while ranking functions are implemented in the score method of BM25TermScorer 
and BM25FTermScorer.

BM25BooleanScorer will create BM25TermScorer or BM25FTermScorer instances depending on the invoked constructor, as next: 

\begin{itemize}
\item To use BM25 ranking function\\
\begin{verbatim}public BM25BooleanQuery(String query, String field, Analyzer analyzer) 
	throws ParseException, IOException\end{verbatim}
\item To use BM25F ranking function\\
\begin{verbatim}public BM25BooleanQuery(String query, String[] fields, Analyzer analyzer) 
	throws ParseException,IOException\end{verbatim}
\end{itemize}

BM25BooleanScorer will ignore any information related to fields that is treated by Lucene QueryParser, thus the search will be carried out only with the field(s), passed as parameters in the constructor. Besides only boolean queries are supported, 
any other query type will be split into terms and executed as a boolean query.

It should be noted that both ranking functions do not use query weights, therefore all computation can be done at scorer level.

\subsubsection*{Scoring}
\begin{itemize}
\item Almost all necessary information in order to compute BM25 relevance can be obtained through the Lucene expert API 
(termdocs, numdocs, docfreq,...), apart from the document average length that can not be obtained directly from the API 
supplied. This value, can be obtained at index time, implementing a specific Similarity that counts and store the length 
of the document fields. As next 

\begin{verbatim}
public class CollectionSimilarityIndexer extends DefaultSimilarity {

  private static Map< String,Long> length = new HashMap<String, Long>();

  @Override
  public float lengthNorm(String fieldName, int numTokens) {
    Long aux = CollectionSimilarityIndexer.length.get(fieldName);
    if (aux==null)
      aux = new Long(0);
    aux+=numTokens;
    CollectionSimilarityIndexer.length.put(fieldName,aux);
    return super.lengthNorm(fieldName, numTokens);
  }
  public static long getLength(String field){
    return CollectionSimilarityIndexer.length.get(field);
  }
}
\end{verbatim}
	
After the indexing process we can retrieve the length of a specific field, and following can be divided by collection numdocs and save the computed value to a file. This value can be 
read when a Searcher is opened. In the provided implementation a method load(String filePath) is supplied in BM25Parameters in order to load average lengths, more details about the file format can be found in the javadoc documentation at \cite{jperezi08b}.

\item The specific BM25 parameters are fixed within the BM25Parameters class, where by default are set at $ k_1 = 2$ and $ b = 0.75$. The BM25F case is more complex, since it needs more specific parameters, mainly an array of string that includes the fields where the term should be searched. All the parameters can be found at 
BM25FParameters, the same $k_1$ is applied. Related to $b$ is set to 0.75 for each field, but is recommended to use better parameters (supplied as a float array) that can be set when the Query 
is initialised. Fixing boost for each field is carried out in a similar fashion, these have been initialised with a value of 1, but it may be supplied with a float array. All BM25F based arrays parameters as $boost_{field}$  and $b_{field}$ must be supplied ordered, that means that for field $i$ into the array of fields, the $boost$ and the $b$ 
parameter for that field will be at $i$ position in both arrays.
\item In both models IDF is computed in BM25Similarity and must be calculated at document level with \textit{docFreq} and \textit{numdocs}. Lucene returns docFreq at field level,that is the number of fields (within documents) where a term $t$ appears. This functionality is not a problem for BM25 since the search is accomplished just in a field. For the BM25F case this is a serious problem, because IDF can not be computed at document level, unless a new field that contains all terms is indexed. The supplied implementation (as an heuristic) computes \textit{docFreq} in the field with the longest average length.
\end{itemize}

\subsection*{How to use it}
The supplied implementation can be used in a similar way as searches are carried out with Lucene, except that \textbf{BM25Parameters or BM25FParameters must be set 
before the query is executed, this has to be done in order to set the average length(s)}, other parameters can be omitted since they are set to default values.

Examples of the BM25 and BM25F raking function appear below:

\subsubsection*{BM25}
\begin{verbatim}
	IndexSearcher searcher = new IndexSearcher("IndexPath");

	//Load average length
	BM25Parameters.load(avgLengthPath);
	BM25BooleanQuery query = new BM25BooleanQuery("This is my Query", 
		"Search-Field",
		new StandardAnalyzer());
	
	TopDocs top = searcher.search(query, null, 10);
	ScoreDoc[] docs = top.scoreDocs;
	
	//Print results
	for (int i = 0; i $<$ top.scoreDocs.length; i++) {
	      System.out.println(docs[i].doc + ":"+docs[i].score);
	}	
\end{verbatim}

\subsubsection*{BM25F}

\begin{verbatim}
	String[] fields ={"FIELD1","FIELD2"};
	IndexSearcher searcher = new IndexSearcher("IndexPath");

	//Set explicit average Length for each field
	BM25FParameters.setAverageLength("FIELD1", 123.5f);
	BM25FParameters.setAverageLength("FIELD2", 42.2f);
	
	//Set explicit k1 parameter
	BM25FParameters.setK1(1.2f);
	
	//Using boost and b defaults parameters
	BM25BooleanQuery queryF = new BM25BooleanQuery("This is my query",
		fields, new StandardAnalyzer());
	
	//Retrieving NOT normalized scorer values
	TopDocs top = searcher.search(queryF, null, 10);
	ScoreDoc[] docs = top.scoreDocs;
	
	//Print results
	for (int i = 0; i $<$ top.scoreDocs.length; i++) {
	      System.out.println(docs[i].doc + ":"+docs[i].score);
	}
\end{verbatim}



\section*{Acknowledgement}
Authors want to thank Hugo Zaragoza for his review and comments.

\end{document}